\documentclass[twoside]{ilcws08}
\usepackage[latin1]{inputenc}
\usepackage[dvipdfm]{graphicx,epsfig,color}
\usepackage{wrapfig,rotating}
\usepackage{amssymb,amsmath,array}

\begin{document}
\title{Tau-pair Performance in ILD Detectors}
\author{Taikan Suehara$^1$, Akiya Miyamoto$^2$, Keisuke Fujii$^2$, \\
Nobuchika Okada$^2$, Hideo Ito$^3$, \\ 
Katsumasa Ikematsu$^2$ and Ryo Yonamine$^4$
\vspace{.3cm}\\
1- The University of Tokyo - International Center for Elementary Particle Physics (ICEPP) \\
7-3-1 Hongo, Bunkyo-ku, Tokyo, 113-0033, Japan
\vspace{.1cm}\\
2- High Energy Accelerator Research Organization (KEK) \\
- Institute of Particle and Nuclear Studies (IPNS) \\
1-1 Oho, Tsukuba, Ibaraki, 305-0801, Japan
\vspace{.1cm}\\
3- The University of Tokyo - Institute for Cosmic Ray Research (ICRR) \\
5-1-5 Kashiwa-no-ha, Kashiwa, Chiba, 277-8582, Japan
\vspace{.1cm}\\
4- The Graduate University for Advanced Studies (SOKENDAI) \\
- School of High Energy Accelerator Science \\
1-1 Oho, Tsukuba, Ibaraki, 305-0801, Japan
}

\maketitle

\begin{abstract}
Tau-pair process has been analyzed in the ILD detector model as a benchmark process for LoI.
Results of background rejection, forward-backward asymmetry and polarization measurements
are obtained with full detector simulation.
Impact of detector parameters for tau-pair analysis is also discussed in this paper.
\end{abstract}

\section{Motivations of tau-pair study}

Tau-pair process (e$^+$e$^-\rightarrow$Z$^\ast\rightarrow\tau^+\tau^-$) at $\sqrt{\mathrm{s}} = 500$ GeV
is one of the benchmark processes\cite{benchmark} proposed by Research Director.
According to the report, this process is a good sample to examine detector performances of
\begin{itemize}
	\item tau reconstruction, aspects of particle flow,
	\item $\pi_0$ reconstruction,
	\item tracking of very close-by tracks.
\end{itemize}
In this process, tau leptons are highly boosted ($\gamma\sim140$), thus decay daughters
(mainly charged and neutral pions, muons and electrons) are concentrated in a very narrow angle. 
Reconstruction of $\pi_0$ from two photons is especially challenging for the ILC detectors,
and much depends on detector parameters, so it is a good measure for detector optimization.

Required observables are cross section, forward-backward asymmetry and polarization of tau leptons.
The polarization measurement requires identification of tau decays, including reconstruction of $\pi_0$.
Efficiency and purity of event selection cuts should also be used for comparison of detector performances.

For physics motivation, tau-pair process is important as a precision measurement of the electroweak theory.
For example, measuring cross section and forward-backward asymmetry of tau-pair process very precisely
can probe existence of heavy Z' boson.

\section{Analysis framework and events}

\subsection{Monte Carlo simulation and detector geometries}

The ILD group has two full detector simulation models, Mokka and Jupiter.
Mokka originates in LDC detector and Jupiter originates in GLD detector,
and both are based on Geant4 Monte Carlo simulation.
In this study I used both simulation models.
Mokka has geometries with detailed implementation of detector components.
I used simulated events processed in Mokka LDCPrime\_02Sc geometry.
The ILD group has simulated quite a large fraction of full Standard Model (SM)
samples, required in the benchmark report, in LDCPrime\_02Sc geometry and 
I used the events to estimate and optimize background suppression.
In contrast, Jupiter has relatively rough geometries and full SM samples have
not been processed, but we have tau-pair samples in several Jupiter geometries 
with three detector parameters, gldapr08\_14m, gldprim\_v04, j4ldc\_v04.
Summary of detector geometries is shown in Fig.~\ref{tab:geometry}.
In rough summary, three geometries in Jupiter differ in sizes,
gldapr08\_14m is large, gldprim\_v04 is middle, j4ldc\_v04 is small in size.
Magnetic field is such that BR for each geometry is almost the same.
LDCPrime\_02Sc is almost as same as gldprim\_v04 in size, but it has
a finer ECAL granularity of 0.5x0.5 cm.
Detailed geometry is much different between Jupiter and Mokka geometries. 

\begin{table}
\centerline{\begin{tabular}{|c|c|c|c|c|}
\hline
Geometry  & gldapr08\_14m & gldprim\_v04 & j4ldc\_v04 & LDCPrime\_02Sc \\\hline\hline
Software  & Jupiter & Jupiter & Jupiter & Mokka  \\\hline
Magnetic field & 3 Tesla & 3.5 Tesla & 4 Tesla & 3.5 Tesla \\\hline
TPC R$_\mathrm{min}$ & 43.7 cm & 43.5 cm & 34.0 cm & 37.1 cm \\\hline
ECAL R$_\mathrm{min}$ & 210 cm & 185 cm & 160 cm & 182.5 cm \\\hline
ECAL thickness & 19.8 cm & 19.8 cm & 19.8 cm & 17.2 cm \\\hline
HCAL thickness & 120 cm & 109 cm & 96 cm & 127.2 cm \\\hline
ECAL granularity & 1x1 cm & 1x1 cm & 1x1 cm & 0.5x0.5 cm \\\hline
\end{tabular}}
\caption{Detector geometries used in this study.}
\label{tab:geometry}
\end{table}

For event reconstruction, including smearing of tracker and calorimeter hits,
tracking, clustering and particle flow, I used MarlinReco framework with
PandoraPFA particle flow algorithm. Raw output of Jupiter is not compatible
with MarlinReco LCIO format, but we have a converter to obtain LCIO files
of Jupiter events. PandoraPFA is especially optimized for Mokka geometries,
so particle flow performance of Jupiter is slightly worse.

\subsection{Event samples}

For signal tau-pair events, we use events generated in DESY.
Whizard 1.51 and TAUOLA are used to generate the events.
SLAC standard samples for LoI are not used because of polarization issues.
I used 80 fb$^{-1}$ signal sample of each geometry
for A$_\mathrm{FB}$ and polarization analysis without background.
For background study, SLAC standard samples are used.
All events simulated in LDCPrime\_02Sc geometry, about 20 million events
in total, are processed with my analysis cuts.

Integrated luminosity is assumed to be 500 fb$^{-1}$ each
for two polarization setups, e$^-_\mathrm{L}$e$^+_\mathrm{R}$
and e$^-_\mathrm{R}$e$^+_\mathrm{L}$.
Assumed polarization ratio is 80\% for electron and 30\% for positron
(i.e.~for e$^-_\mathrm{L}$e$^+_\mathrm{R}$ setup 90\% of electrons are
leftly polarized and 65\% of positrons are rightly polarized).

\subsection{Tau clustering}

For tau clustering, an original clustering processor (TaJet) is applied to
the output of PandoraPFA. Following is a procedure of the processor.
\begin{enumerate}
	\item Sort particles in energy order.
	\item Select the most energetic charged particle (a tau candidate).
	\item Search particles to be associated to the tau candidate. Criteria is:
	\begin{enumerate}
		\item Opening angle to the tau candidate is smaller than 50 mrad., or
		\item Opening angle to the tau candidate is not larger than 1 rad.~and invariant mass with the tau candidate is less than 2 GeV (m$_\tau$ = 1.777 GeV). 
	\end{enumerate}
	\item Combine energy and momentum of the tau candidate and associated particle and treat the combined particle as the new tau candidate.
	\item Repeat from 3.
	\item After all remaining particles do not meet the criteria, remaining most energetic charged particle is the next tau candidate. (Repeat from 2.)
	\item After all charged particles are associated to tau candidates, remaining neutral particles are independently included in the cluster list as neutral fragments.
\end{enumerate}

In the clustering stage, events with $>$ 6 tracks are pre-cut to accelerate clustering since $>$ 99\% of tau decays have $\leq$ 3 charged particles.
Event with only one positive and one negative tau clusters are processed with latter analysis.

\section{Background suppression}

Main background of tau-pair analysis is Bhabha (e$^+$e$^-\rightarrow$e$^+$e$^-$), WW $\rightarrow$ $\ell\nu\ell\nu$ and $\gamma\gamma\rightarrow\tau^+\tau^-$.
Since cross sections of Bhabha and two-photon events are huge (about $10^4$ and $10^3$ larger than signal, respectively),
we need tight selection cuts for those background events.
Following cuts are applied to signals and all SM background events after the tau clustering.

\begin{enumerate}
	\item Number of tracks $\leq$ 6. Included as a pre-cut in tau clustering processor.
	\item Only one positive and one negative tau clusters must exist in the event.
	\item Opening angle of two tau candidates must be $>$ 178 deg.

		This cut efficiently suppresses WW $\rightarrow$ $\ell\nu\ell\nu$ background.
	\item ee and $\mu\mu$ events are rejected.

		Charged particles depositing $>$ 90\% of their energy in ECAL are identified as electrons, and
		charged particles depositing $<$ 70\% of their energy (estimated by curvature of their tracks) in ECAL+HCAL are identified as muons.
		Events with two electrons or two muons are rejected in this cut.
		This cut is especially for suppressing Bhabha and e$^+$e$^-\rightarrow\mu^+\mu^-$ events. Signal loss is about 6\%.
	\item $|\cos\theta| < 0.9$ for both tau clusters.

		t-channel Bhabha events are almost completely suppressed by this cut. 20\% of signal events are lost.
	\item $40 < \mathrm{E}_\mathrm{vis} < 450$ GeV.

		Lower bound suppresses $\gamma\gamma\rightarrow\tau^+\tau^-$ events, and upper bound suppresses Bhabha events. Signal lost is negligibly small.
\end{enumerate}

\begin{wraptable}{r}{0.6\columnwidth}
\centerline{\begin{tabular}{|c|c|c|}
\hline
Cuts & Signal & Background \\ \hline\hline
\# tracks, \# clusters          & $5.7\times10^5$ & $7.9\times10^8$ \\\hline
Opening angle $>$ 178 deg.      & $1.6\times10^5$ & $1.3\times10^8$ \\\hline
$|\cos\theta| < 0.9$            & $1.3\times10^5$ & $1.2\times10^7$ \\\hline
ee, $\mu\mu$ veto               & $1.2\times10^5$ & $6.2\times10^5$ \\\hline
45 $<$ E$_\mathrm{vis}$ $<$ 450 & $1.2\times10^5$ & $1.3\times10^4$ \\\hline
\end{tabular}}
\caption{Cut statistics for background suppression.}
\label{tab:smcuts}
\end{wraptable}

Table \ref{tab:smcuts} shows the result of these cuts. $\gamma\gamma\rightarrow\tau^+\tau^-$ background is currently not included, but
generator-level study shows the effect of the $\gamma\gamma\rightarrow\tau^+\tau^-$ background is not significant (most events are eliminated
by the E$_\mathrm{vis}$ cut). Number of events are normalized to 500 fb$^{-1}$.
e$^-_\mathrm{L}$e$^+_\mathrm{R}$ polarization (80\% and 30\%, respectively) is assumed.
In total, number of background events is about 10\% of signal, not significant.
If we assume that we know the shape of the background, statistical error of the background is much smaller than signal statistics
and negligible for the further study. 
$\gamma\gamma\rightarrow\tau^+\tau^-$ background is planned to be included in the LoI study.

\section{Forward-backward asymmetry}

\begin{wrapfigure}{r}{0.5\columnwidth}
\centerline{
	\includegraphics[width=0.45\columnwidth]{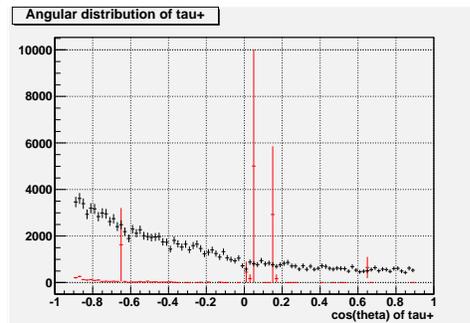}
}
\caption{Angular distribution of $\tau^+$ momentum direction. Black crosses show signal and red crosses show background.
Vertical axis is normalized to 500 fb$^{-1}$. Error bars represent errors in current MC statistics.
}
\label{fig:afb}
\end{wrapfigure}

Figure \ref{fig:afb} shows a simulated result on angular distribution of $\tau^+$ leptons in LDCPrime\_02Sc model.
All events passed criteria described in the previous section are put into the histograms.
Clear asymmetry can be seen in signal distribution.
There are several bins where background is very large, but this is a result of low MC statistics of Bhabha, $\gamma^\ast\gamma^\ast$
and $e\gamma^\ast$ processes (about 0.1 fb$^{-1}$ each).
We plan to improve MC statistics of those events by applying preselections at generator level.

Forward-backward asymmetry can be calculated by
\begin{eqnarray}
	A_{FB} &=& \frac{N_F - N_B}{N_F + N_B}, \\
	\delta{}A_{FB} &=& \frac{2\sqrt{N_FN_B(N_F+N_B)}}{(N_F+N_B)^2},
\end{eqnarray}
where $N_B$ is number of events in backward region ($\cos\theta<0$) and $N_F$ is number of events in forward region ($\cos\theta>0$).
By estimated number of signal events ($N_F=8956, N_B=2893$), $A_{FB}$ is estimated to be 51.2$\pm$0.25\%
(background statistics is not included).
Since total estimated number of background events is about 10\% of signal, effect of background to statistical error is
smaller than signal statistics if background distribution can be well determined.

\section{Polarization analysis}
\subsection{Event selection}

There are five dominant decay modes of tau leptons, $\tau^+ \rightarrow \mathrm{e}^+\overline{\nu_\mathrm{e}}\nu_\tau$ (17.9\%),
$\tau^+ \rightarrow \mu^+\overline{\nu_\mu}\nu_\tau$ (17.4\%),
$\tau^+ \rightarrow \pi^+\nu_\tau$ (10.9\%),
$\tau^+ \rightarrow \rho^+\nu_\tau \rightarrow \pi^+\pi^0\nu_\tau$ (25.2\%),
and $\tau^+ \rightarrow \mathrm{a}_1^+\nu_\tau \rightarrow \pi^+\pi^0\pi^0\nu_\tau$ (9.3\%).

Since first two modes are leptonic 3-body decay and polarization information is partially lost due to the missing neutrinos,
and the last a$_1$ mode has relatively low branching ratio,
we currently use only $\tau^+ \rightarrow \pi^+\nu_\tau$ and $\tau^+ \rightarrow \rho^+\nu_\tau$ modes.
These modes are selected by following criteria.

\begin{enumerate}
	\item Tau clusters with one charged tracks are selected.
	\item Clusters with electrons and muons are eliminated. Muon identification is the same as that in SM suppression cut.
		Electron identification is ECAL deposit $>$ 90\% for $\tau^+ \rightarrow \pi^+\nu_\tau$ selection
		and $>$ 97\% for $\tau^+ \rightarrow \rho^+\nu_\tau$ selection.
	\item Clustered with energy $>$ 10 GeV is eliminated (since lepton identification is currently poor in low energy clusters).
	\item Check whether neutral particles are associated in the cluster.
		If $>$ 1 GeV neutral particles are not associated, the cluster is treated as a $\tau^+ \rightarrow \pi^+\nu_\tau$ event.
		If $>$ 10 GeV neutral particles are associated, the cluster is treated as a $\tau^+ \rightarrow \rho^+\nu_\tau$ event candidate.
	\item For $\tau^+ \rightarrow \rho^+\nu_\tau$ candidates, invariant mass of $\rho$ is calculated from 4-momenta of charged pion
		and whole cluster. Clusters with invariant mass around 200 MeV to m$_\rho$ (570 to 970 MeV) are accepted.
	\item Optional $\pi^0$ mass cut is applied to the $\tau^+ \rightarrow \rho^+\nu_\tau$ candidates.
		In this cut, invariant mass of $\pi^0$ is calculated with clusters which have $\geq$ 2 neutral particles 
		Events with invariant mass of 0 to 200 MeV (m$_\pi^0$ = 135 MeV) are accepted.
		All events with only one neutral particle are eliminated by this cut (if applied).
\end{enumerate}

\begin{figure}
	\begin{minipage}[t]{.47\textwidth}
		\includegraphics[width=0.95\columnwidth]{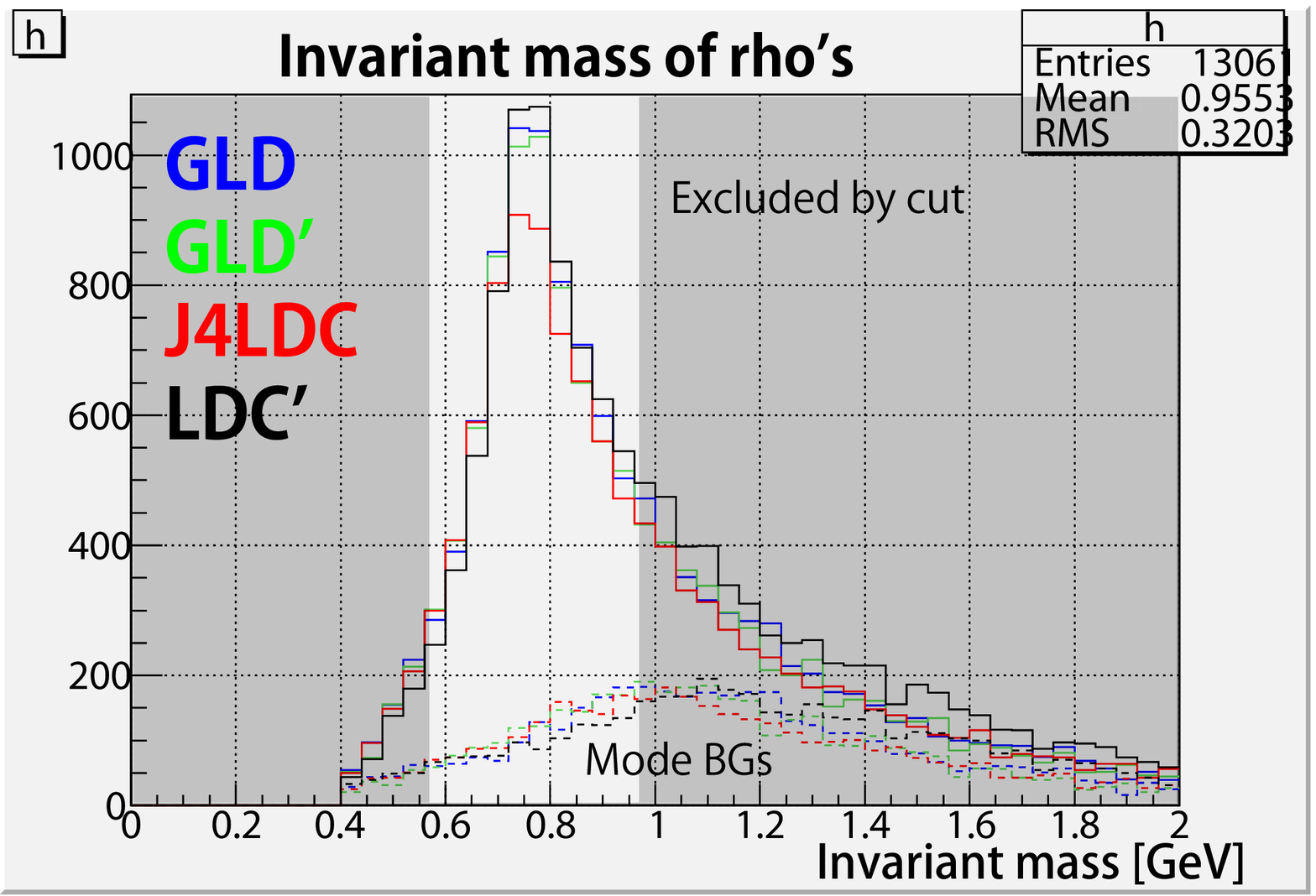}
		\caption{Invariant mass distribution of $\rho$ with four geometries.}
		\label{fig:rhomass}
	\end{minipage}
\hfill
	\begin{minipage}[t]{.47\textwidth}
		\includegraphics[width=0.95\columnwidth]{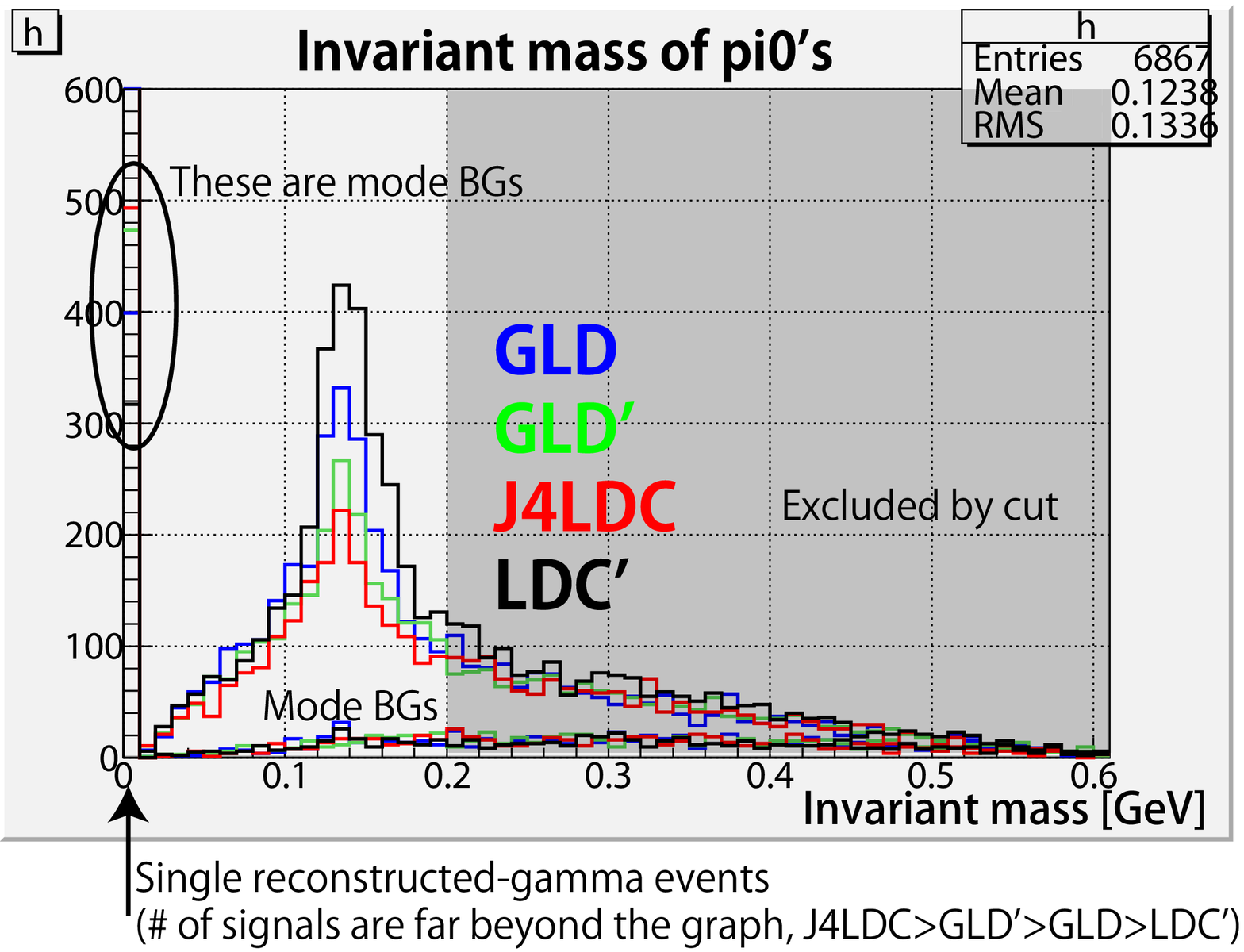}
		\caption{Invariant mass distribution of $\pi^0$ with four geometries.}
		\label{fig:pi0mass}
	\end{minipage}
\end{figure}

Figure \ref{fig:rhomass} and \ref{fig:pi0mass} shows invariant mass distribution of $\rho$ and $\pi^0$ with
gldapr08\_14m (noted GLD in the plot), gldprim\_v04 (GLD'), j4ldc\_v04 (J4LDC) and LDCPrime\_02Sc (LDC') geometries.
Especially $\pi^0$ invariant mass distribution shows large difference between geometries.
Larger and higher granularity geometry apparently gives better results in $\pi^0$ reconstruction.

\begin{table}
\centerline{\begin{tabular}{|c|c|c|c|c|}
\hline
Geometry                         & gldapr08\_14m & gldprim\_v04 & j4ldc\_v04 & LDCPrime\_02Sc \\\hline\hline
$\pi$ mode efficiency            & 21.3\%        & 21.4\%       & 21.4\%     & 21.2\% \\\hline
$\pi$ mode purity                & 85.7\%        & 83.6\%       & 80.8\%     & 88.5\% \\\hline
$\rho$ mode eff. wo/ $\pi^0$ cut & 12.7\%        & 12.1\%       & 11.3\%     & 12.8\% \\\hline
$\rho$ mode pur. wo/ $\pi^0$ cut & 83.4\%        & 81.8\%       & 81.4\%     & 85.7\% \\\hline
$\rho$ mode eff. w/ $\pi^0$ cut  & 5.31\%        & 4.32\%       & 3.72\%     & 6.38\% \\\hline
$\rho$ mode pur. w/ $\pi^0$ cut  & 92.3\%        & 90.3\%       & 90.5\%     & 93.9\% \\\hline
\end{tabular}}
\caption{Selection efficiency and purity for polarization analysis.}
\label{tab:polcuts}
\end{table}

Table \ref{tab:polcuts} shows a result of selection efficiency and purity for each detector geometry.
For $\tau^+ \rightarrow \rho^+\nu_\tau$ mode, both results of selection without $\pi^0$ invariant mass cut and
selection with $\pi^0$ invariant mass cut are shown in parallel. $\pi^0$ mass cut gives higher purity in
event selection but efficiency becomes much less, thus the cut seems not practical for analysis in current geometries.
For all selection LDCPrime\_02Sc gives the best result, and gldapr08\_14m follows the next.

\subsection{Polarization measurement}

\begin{figure}
	\begin{minipage}[t]{.47\textwidth}
		\includegraphics[width=0.95\columnwidth]{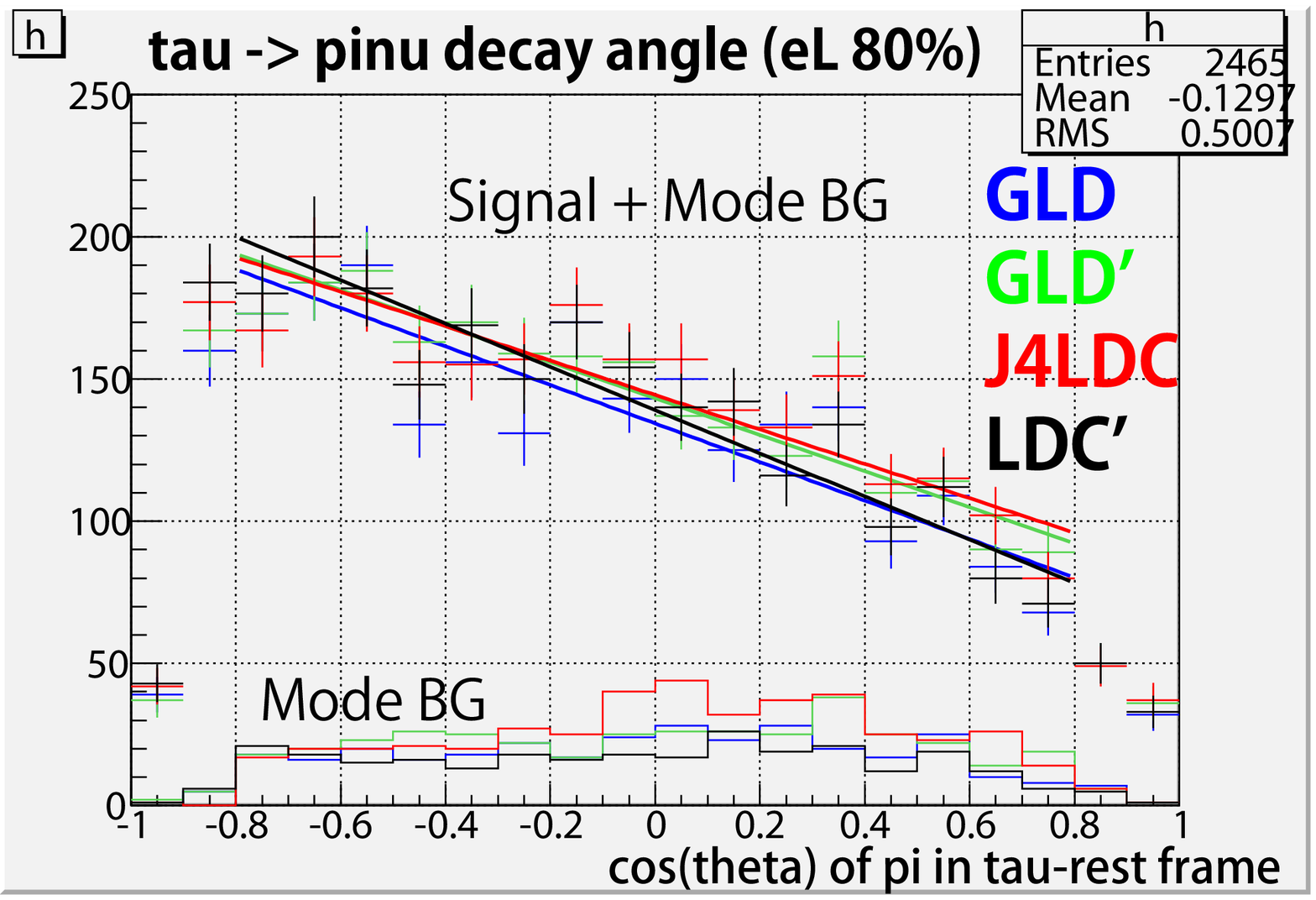}
	\end{minipage}
\hfill
	\begin{minipage}[t]{.47\textwidth}
		\includegraphics[width=0.95\columnwidth]{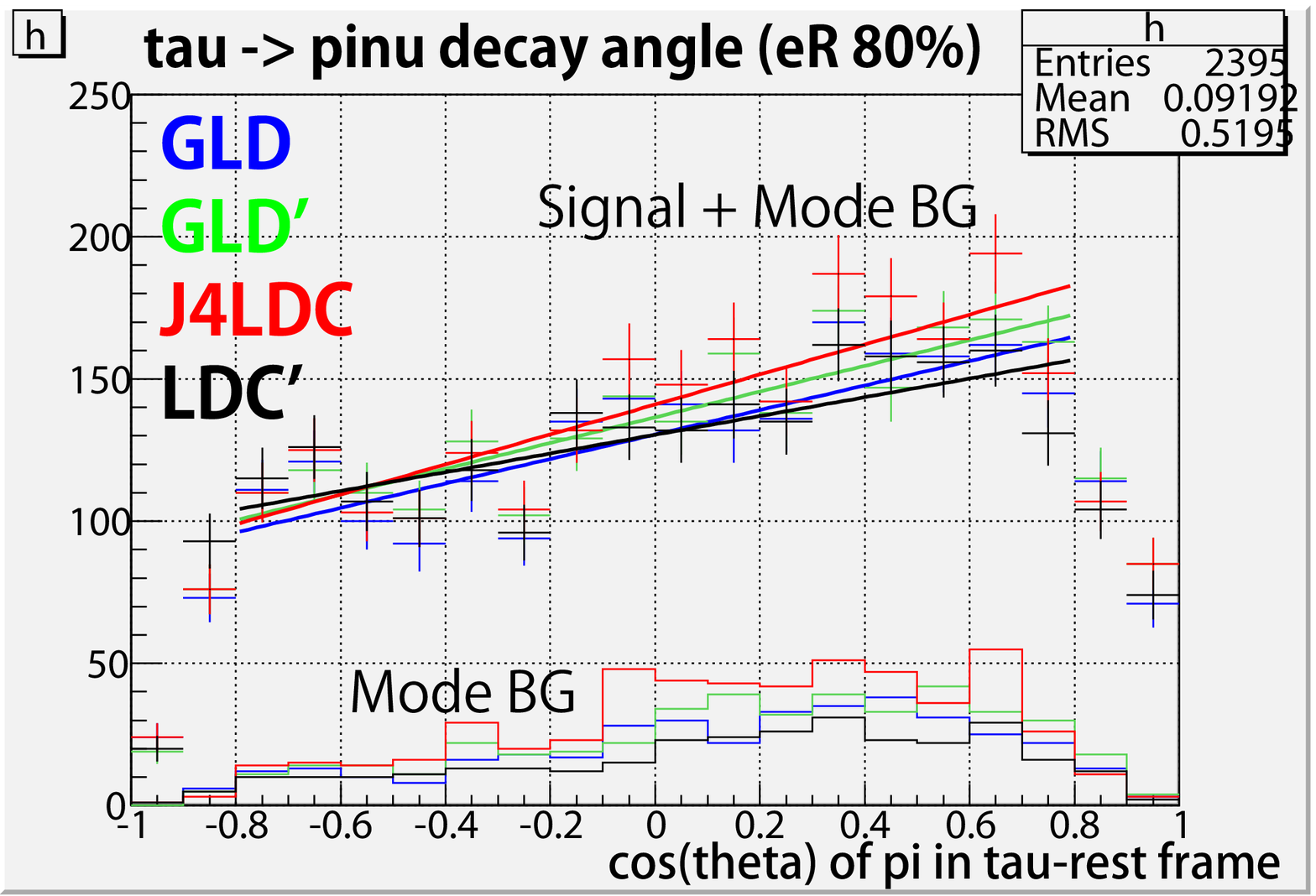}
	\end{minipage}
	\caption{Angular distribution of charged pion in tau-momentum frame,
		left: e$^-_\mathrm{L}$e$^+_\mathrm{R}$ polarization, right: e$^-_\mathrm{R}$e$^+_\mathrm{L}$ polarization.}
	\label{fig:pitheta}
\end{figure}

To obtain polarization of $\tau^+ \rightarrow \pi^+\nu_\tau$ events,
cos$\theta$ distribution of pion momentum direction with respect to tau momentum direction should be observed.

Figure \ref{fig:pitheta} shows the cos$\theta$ distribution.
For the e$^-_\mathrm{L}$e$^+_\mathrm{R}$, number of events is larger in $\cos\theta < 0$ area, and
for the e$^-_\mathrm{R}$e$^+_\mathrm{L}$, number of events is larger in $\cos\theta > 0$ area.
Polarization of tau leptons can be determined by the ratio of number of events between left and right half
of the graph, or by linear fit of the histograms.
Analysis shows the polarization can be determined by 1.2-1.3\% statistical error,
but the amount of remaining background varies by geometries as a result of difference in selection purity
(for detailed numbers, see the slide\cite{url}).

\begin{figure}
	\begin{minipage}[t]{.47\textwidth}
		\includegraphics[width=0.95\columnwidth]{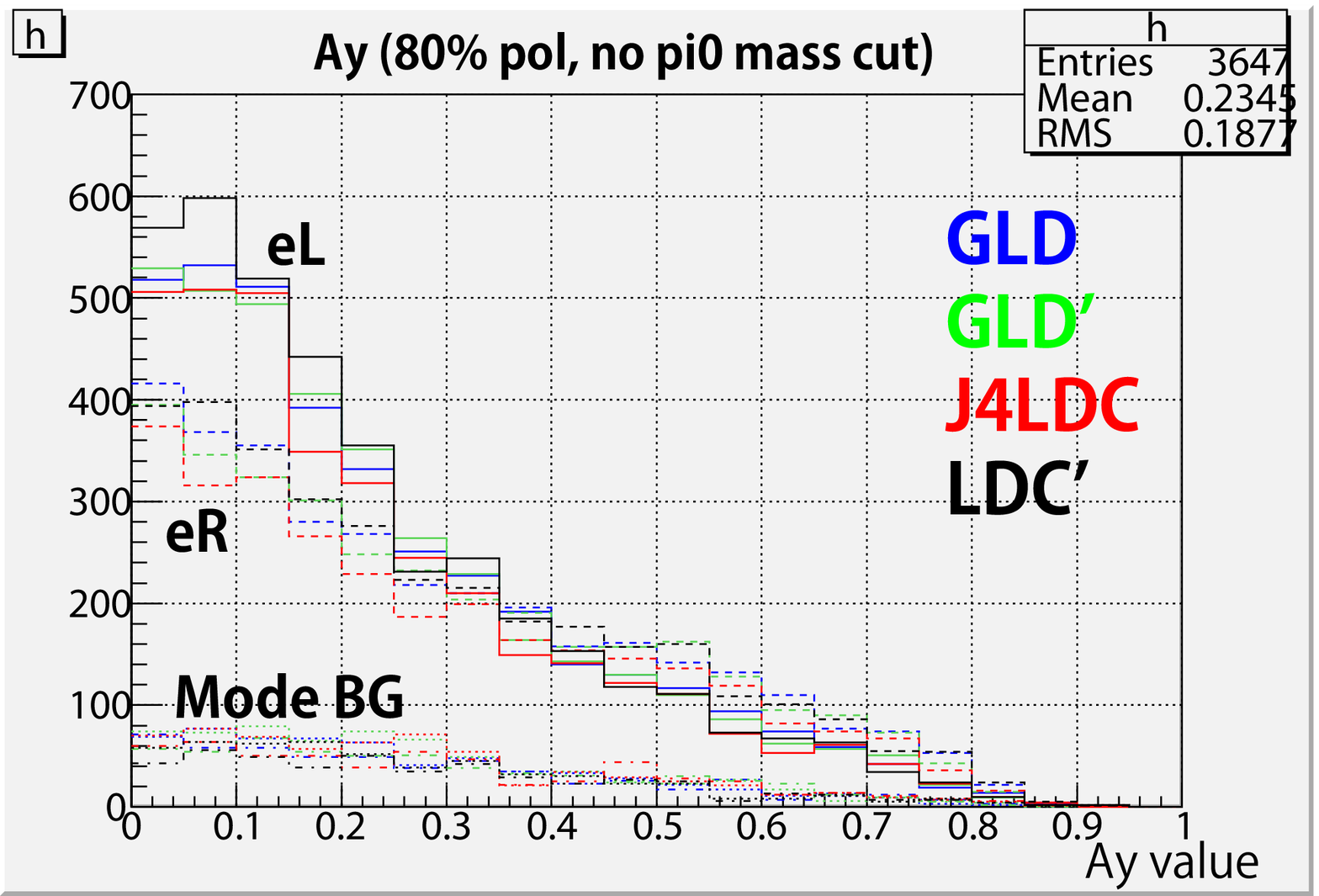}
	\end{minipage}
\hfill
	\begin{minipage}[t]{.47\textwidth}
		\includegraphics[width=0.95\columnwidth]{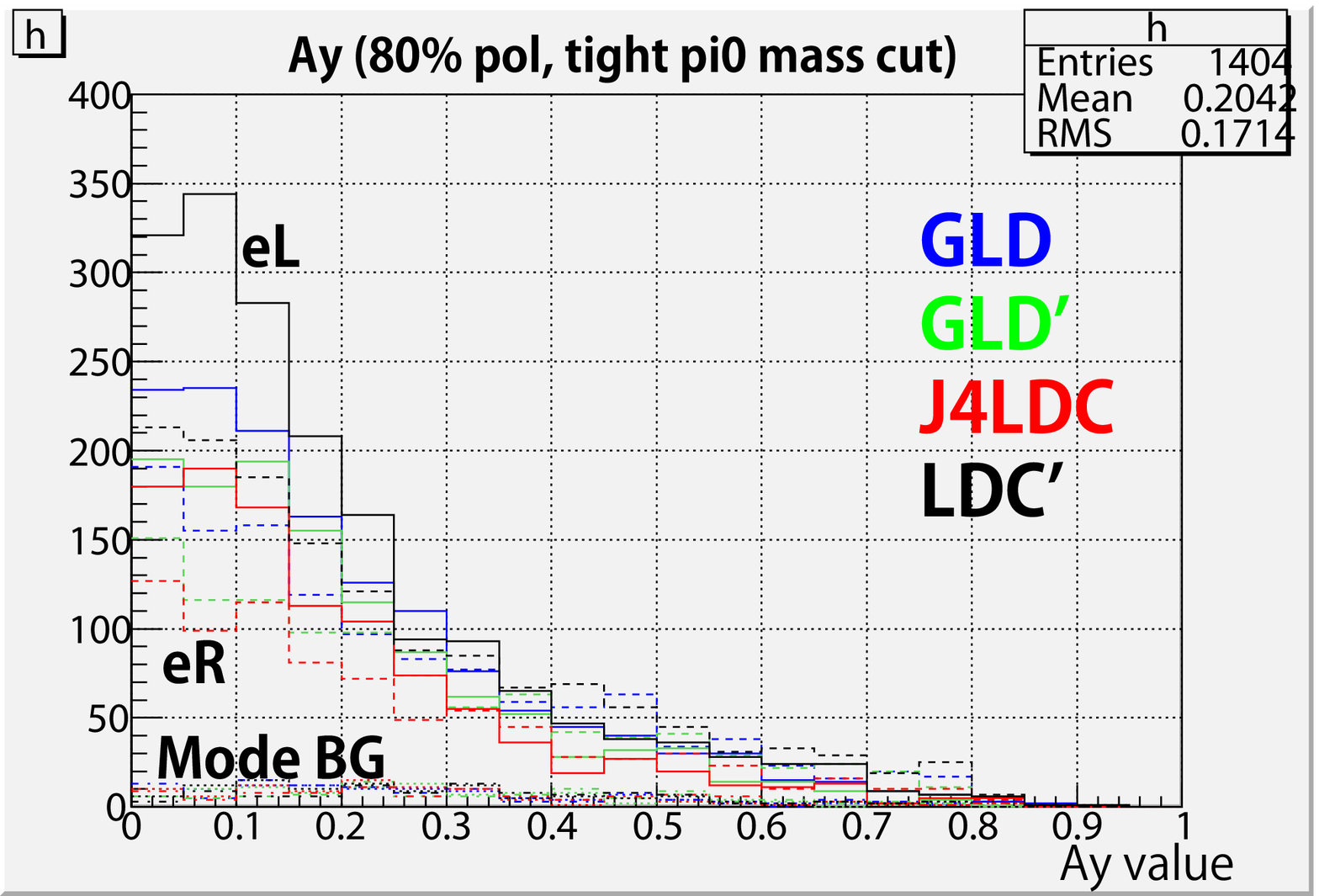}
	\end{minipage}
	\caption{Distribution of parameter $y$. See text for details of $y$. left: mode selection without $\pi^0$ invariant mass cut, 
		right: selection with $\pi^0$ mass cut. }
	\label{fig:rhopol}
\end{figure}

For analysis of $\tau^+ \rightarrow \rho^+\nu_\tau$ polarization, analysis is more complicated.
To obtain polarization of tau leptons, we can use $\rho$ polarization, indicated by angular distribution of
$\rho^+ \leftarrow \pi^+\pi^0$ decay, in addition to $\rho$ angular distribution with respect to $\tau$ frame.
To combine those indicator, we use $y$ parameter, defined in \cite{rhopol} as
\begin{equation}
	y = \frac{E_{\pi^0} - E_{\pi^+}}{E_\tau}.
\end{equation}
According to \cite{rhopol}, polarization of tau leptons can be determined by 
\begin{equation}
	0.85P_\tau = \frac{N(y>y_c)}{N(y>y_c;P_\tau=0)} - \frac{N(y<y_c)}{N(y<y_c;P_\tau=0)}, y_c=0.316.
\end{equation}

Figure \ref{fig:rhopol} shows the $y$ distribution. From the distribution, $P_\tau$ can be determined
by 1.1-1.2\% statistical error if we do not apply $\pi^0$ mass cut, and 1.7-2.3\% with $\pi^0$ mass cut
(again, detailed number can be seen in the slide\cite{url}).
Difference can be observed between geometries, due to selection efficiency, but
with no $\pi^0$ mass cut the difference is not so large.

\section{Summary}

Tau-pair process has been analyzed in the ILD detector models.
It is found that tau-pair forward-backward asymmetry observation of better than 1\% resolution
can be achieved in the current ILD detector models and no large difference between
detector models are found.
For the polarization analysis, clear dependence is seen in $\pi_0$ reconstruction.
LDCPrime\_02Sc gives the best result and gldapr08\_14m follows.
With the result we can estimate that larger and more granular detector models give better results.


\begin{footnotesize}



%

\end{footnotesize}


\end{document}